# Beyond single-crystalline metals: ultralow-loss silver films on lattice-mismatched substrates

Aleksandr S. Baburin[1,2], Dmitriy O. Moskalev[1], Evgeniy S. Lotkov[1], Olga S. Sorokina[1], Dmitriy A. Baklykov[1], Sergey S. Avdeev[1], Kirill A. Buzaverov[1], Georgiy M. Yankovskii[2], Alexander V. Baryshev[2], Ilya A. Ryzhikov[1] and Ilya A. Rodionov[1,2,*]

[1]FMN Laboratory, Bauman Moscow State Technical University, Moscow 105005, Russia
[2]Dukhov Automatics Research Institute, (VNIIA), Moscow 127055, Russia
*Correspondence: irodionov@bmstu.ru

## ABSTRACT

High-quality factor plasmonic devices are crucial components in the fields of nanophotonics, quantum computing and sensing. The majority of these devices are required to be fabricated on non-lattice matched or transparent amorphous substrates. Plasmonic devices quality factor is mainly defined by ohmic losses, scattering losses at grain boundaries, and in-plane plasmonic scattering losses of a metal – substrate system. Here, we demonstrate the deposition technique to e-beam evaporate ultralow-loss silver thin films on transparent lattice-mismatched substrates. The process is based on evolutionary selection growth. The key feature of our approach is a well-defined control of deposition on a cooled substrate, self-crystallization and subsequent annealing for precise stress relaxation that promote further grains growth. We are able to deposit 100-nm thick ultraflat polycrystalline silver films with micrometer-scale grains and ultralow optical losses. Finally, we show ultra-high-quality factor plasmonic silver nanostructures on transparent lattice-mismatched substrate comparable to epitaxial silver. This can be of the great interest for high performance or single-molecule optical sensorics applications.

## Introduction

A number of novel devices in quantum science [1-4], nanophotonics [5, 6], plasmonics [7, 8] and sensing [9-11] require simple methods to deposit metal films with state-of-the-art performance and microstructure. It becomes the dominant factor in the fields like plasmonics, which is used in different applications, including SERS [9-11], strong light confinement waveguides [12-14], diagnostics and treatment of cancer [15, 16], ultrasensitive detection [17 - 22]. Proper materials selection is essential to enable high performance for all the types of plasmonic devices. The best plasmonic films have the minimal permittivity imaginary part $\varepsilon''$ and the maximal absolute value of the real part $\varepsilon_1$ [8, 23-26]. Plasmonic devices performance is usually limited by "metal thin film on substrate" criterion. For example, the majority of sensors are fabricated using transparent substrates [19, 20, 22, 27, 28] and waveguides require multi-layered low-loss film stack deposition [12-14]. Silver (Ag) shows the outstanding performance among all the plasmonic materials [8, 25-35]. It has also been experimentally confirmed that single-crystalline Ag films have the best properties [8, 32, 33, 35]. However, there is still no common approach to deposit high-quality polycrystalline silver films on lattice-mismatched substrates [30, 32, 34-38]. Usually, polycrystalline films show noticeably inferior optical properties due to the large number of grain boundaries, imperfect structure, rough surface and defective interfaces compared to single crystalline [8, 30, 32, 33, 37-39]. Moreover, in the case of plasmonic devices based on polycrystalline metal films (with the typical grain size ~10-100 nm [31, 34, 35]) it's difficult to fabricate precise nanostructures using nanolithography techniques [40]. Therefore, deposition of polycrystalline films with bigger grains is a powerful way for plasmonic devices performance improvement. At the same time, it is important to avoid film roughness increasing during thin film grains growth. That is why the best results for polycrystalline films were obtained using the film stripping technology [36, 39]. However, scaling up is the well-known challenge for this method. In this report, we describe a scalable technology to synthesize an ultralow-loss silver polycrystalline big grains (PCBG) thin films on lattice-mismatched substrates for subsequent nanoscale patterning of plasmonic devices.

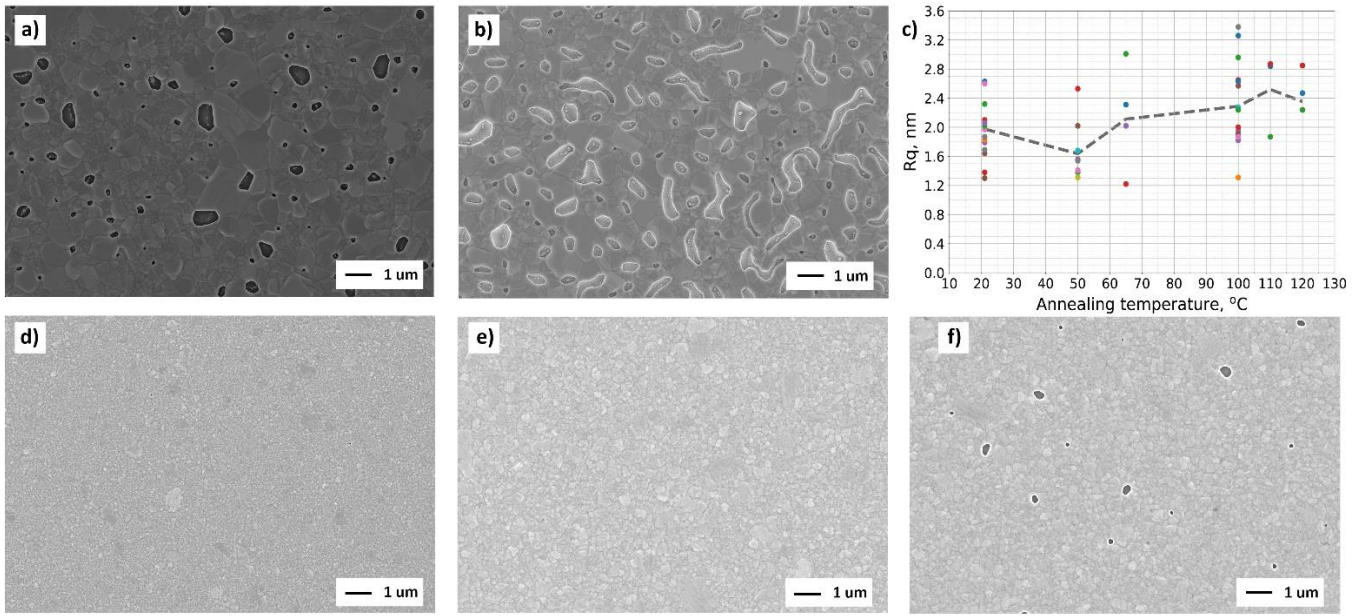

**Figure 1.** Silver (100 nm) thin film SEM images (7000x magnification), deposited on: (a) quartz; (b) sapphire; (c) film root mean square roughness on annealing temperature; silver (100 nm) thin film SEM image (7000x magnification), deposited on quartz, annealed at (d) 150°C; (e) 300°C; (f) 500°C.

In general, increasing the substrate temperature and decreasing deposition rate is known to lead to the formation of a film with big grains [35, 36]. However, this approach does not take into account the dewetting effect, which is especially well-observed for the Ag films. Its influence is mostly apparent for the deposition at substrate temperatures above 100°C on lattice-mismatched substrates [29, 41, 42]. Films do not overcome a coalescence barrier and remain noncontinuous even at thicknesses higher than hundred nanometers (Fig. 1a, b). It also should be noted that film grain growth after deposition has an even higher influence on the final structure. One of the possible ways to further increase grain size is film post-annealing [43]. However, annealing of Ag films deposited at room temperature on lattice-mismatched substrates does not improve its structural and optical properties [43–45]. During heat treatment, the film roughness grows and then the film breaks (Fig. 1c-f) due to poor wetting of the surface. Therefore, standard techniques do not provide the desired Ag thin film structure. It is reported that the common thin film growth theory is not yet finalized [46]. Consequently, a deeper study of polycrystalline films growth process is required. We reviewed factors affecting the film growth (See Section Results and Discussion) to identify the necessary conditions for the implementation of the evolutionary selection method to obtain films with big grains using the three-step deposition process. Proposed process consists of following steps (Fig. 2a, b): 1) deposition of nanocrystalline film with sufficient stored internal energy, 2) self-crystallization and stored energy relaxation, 3) final crystallization at well-tuned annealing.

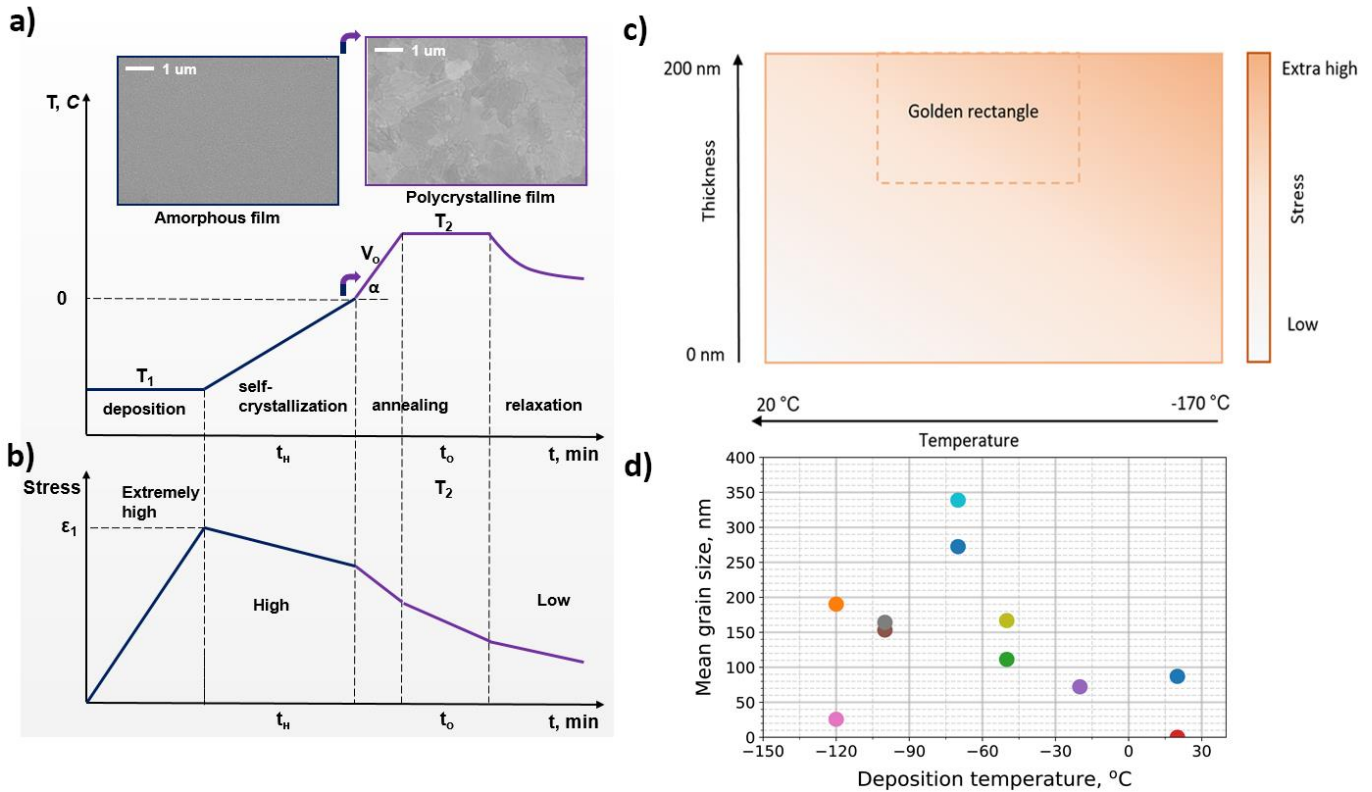

**Figure 2.** a) Temperature evolution during the deposition of Ag films; b) internal stresses evolution during the deposition of Ag films; c) thermal map of stress, film thickness and deposition temperature (the rectangle marks the region with the thickness/temperature ratio, which provides the maximum grain size after recrystallization); d) film grain size on deposition temperature relation.

**Step 1.** At the deposition stage, the film growth is determined by the ratio of the metal – substrate system energy and the flow of the incoming material atoms. At low (cryogenic) substrate temperatures, a deposited Ag film shows a nanocrystalline structure with a typical grain size ~10-100 nm. The film structure could remain as-deposited for some special cases of crystallization. However, commonly small grains of deposited film have low energy and are ready to coalesce in larger ones providing further crystallization. At this stage, film grain size, the accumulated internal stress, the film thickness and the number of defects (details could be found in Supplementary Section 4) are the key growth-defining parameters. On this step grain size is determined by the deposition rate and substrate temperature. Both the increase of the deposition rate and the decrease of substrate temperature favors the decrease of the grain size [47-51]. Stress, in its turn, is mostly determined by film thickness and deposition temperature. Depending on stress formation mechanisms, there are two types of metals (See Supplementary Fig. S9). Silver deposited at a low-temperature substrate behaves as Type 1 metal. Significant film stress growth is observed for temperatures below –80°C [52]**.** We experimentally found the minimum film thickness (at least 70 nm) required to provide film self-crystallization during subsequent heating up to the room temperature. This is due to the fact that stress reaches minimal required value at the critical thickness [52] following the increase of film's thickness. For each metal – substrate system one can find the thickness/temperature ratio providing the maximum grain size during recrystallization. For the system discussed in this article, the area corresponding to the maximum grain size is marked as "Golden rectangle" in Figure 2c.

**Step 2.** On the second step self-crystallization starts at the passive film heating. The film undergoes stress relaxation that is re-distributed at the microlevel leading to the film crystallization and the grain growth. It could be considered as a self-crystallization, because there are no external forces to promote crystallization. At this point, stress relaxation time is the main optimization parameter. It should be sufficiently long to favor the evolution selection growth becoming dominant. We experimentally found out that the stress relaxation time should exceed 12 hours.

**Step 3.** Thermal post-treatment improves a film structure as it promotes crystallization. During this step, the bigger grains are forming due to the evolutionary selection growth eliminating small grains. At this step the annealing temperature should not exceed the dewetting temperature (for a thin Ag film <100°C). Otherwise, it initializes a three-dimensional growth leading to the increased film roughness [43-45]. The annealing time should be adjusted accordingly to overcome the thermal grooving effect (details could be found in Supplementary Sections 1-2). Films are expected to be heated in vacuum or oxygen-free environment. Thin film should not even be exposed to an oxygen atmosphere between the deposition and the annealing steps. These requirements are caused by the relatively high chemical activity of silver: adsorbed impurities become the points of a three-dimensional crystallization.

**Experimental section.** We deposited a series of 100-nm thick Ag films on quartz substrates at cryogenic temperatures. Films were expected to show a nanocrystalline structure. The films were heated to room temperature after deposition and subsequently annealed. The ratios of the deposition and annealing parameters were chosen to obtain an optimized film structure (grains size noticeably larger than film thickness). To maximize the grain size, we optimized each process step: deposition, self-crystallization and annealing (Table 1).

**Table 1.** Ag thin film deposition parameters.

|  | Series 1 | Series 2 | Series 3 | Series 4 | Series 5 | Series 6 | Series 7 | Series 8 | Series 9 | Series 10 | Series 11 |
|---|---|---|---|---|---|---|---|---|---|---|---|
| Substrate T, °C | 20 | -120 | -50 | 20 | -20 | -100 | -120 | -100 | -50 | -70 | -70 |
| Annealing T, °C | 100 | 120 | 120 | 20 | 120 | 60 | 60 | 60 | 60 | 120 | 120 |
| Heating t, min | 40 | 110 | 100 | - | 20 | Passive +30 | Passive +90 | Passive +30 | Passive +30 | Passive +30 | 30 |
| Heat rate, °C/min | 5 | 5 | 5 | - | 5 | 1 | 1 | 1 | 1 | 5 | 5 |
| dT, C | 80 | 240 | 170 | 0 | 140 | 160 | 180 | 160 | 110 | 190 | 190 |

For the deposition step, we obtained the film grain size dependence on the substrate temperature (Fig. 2d). The function extremum can be clearly seen at the point corresponding to -70°C, which also corresponds to the "Golden rectangle" area maximum stress (Fig. 2c). Another important parameter of the deposited films is the grain fill factor. It is determined as the ratio of film surface covered with the desired grain size to the full film surface. This parameter is even more representative for the further precise structures fabrication with lithography techniques. We studied the impact of the deposition temperature on the fill factor of big grains (>0.5 μm) and small grains (<0.1 μm) as shown in Fig. 3a, d. The film of desired structure should have highest possible fill factor of grain >0.5 um and lowest possible fill factor of grain <0.1 μm.

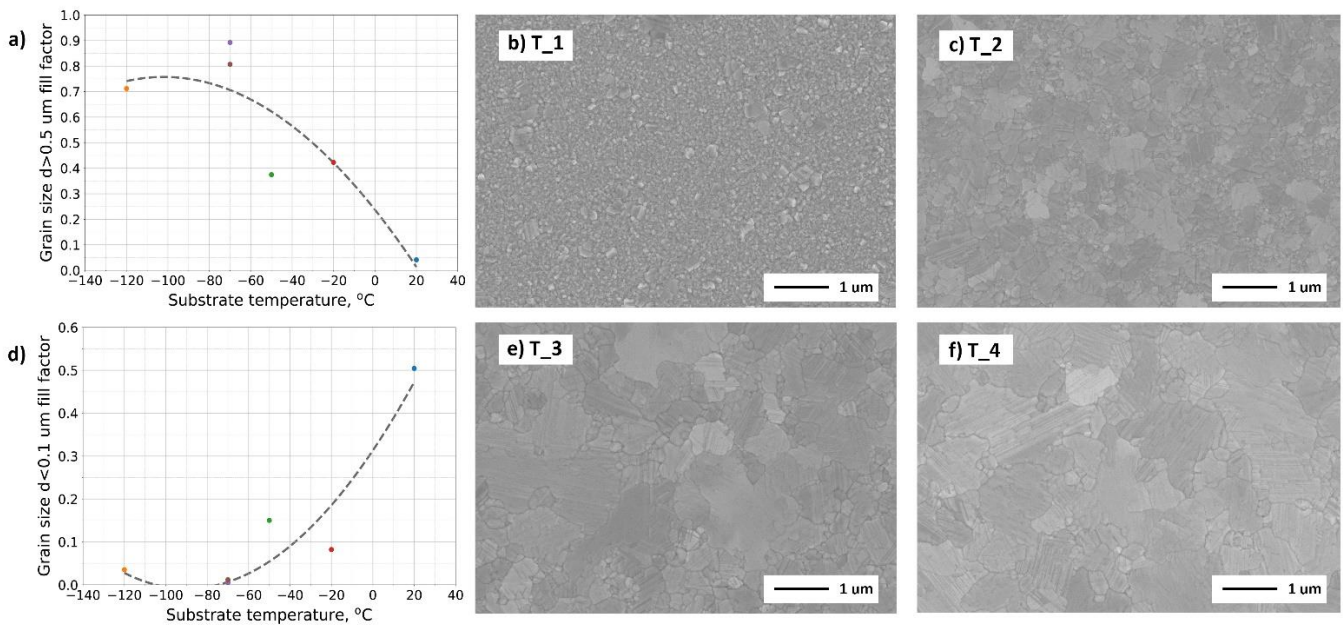

**Figure 3.** Grain size fill factor on deposition temperature for (a) d > 0.5 um, (d) d < 0.1 um (b, c, e, f) SEM images (7000x magnification) of Ag films at different annealing temperatures (temperature rises from T_1 to T_4)

Grain fill factor plots also exhibits extremum points at -70°C substrate temperature: big grains (>0.5 μm) occupy more than 80% of the film surface while the surface, occupied by small grains (<0.1 μm), is less than 1 %. We also obtained dependence of the grain fill factor on the difference between the annealing temperature and the substrate temperature during deposition (ΔT=T2-T1) (See Supplementary Fig. S8). This relation confirms that an increase in the ΔT parameter contributes to an increase in the stress (see equation 4) and provides formation of bigger grains. The examples of the Ag film structure at different annealing temperatures are shown in Figure 3b, c, e, f.

**Results and Discussion.**

We propose an original approach to overcome the fundamental limitation caused by the Ag dewettability following method of evolutionary selection for the thin films growth control [53]. This method, in its turn, is based on choosing the preferred orientation, which would inhibit growth of other crystallographic planes. We adapted this method to the crystallization process of an already deposited film in order to obtain polycrystalline films with large grains.

The necessary conditions for the evolutionary selection growth are following:
a) deposition of the nanocrystalline film compatible with further grain growth,
b) sufficient stored energy in the film to induce further crystallization,
c) sufficient relaxation period of this energy to provide the preferred orientation growth.

The following forces are typically used to describe film crystallization [53, 54].

For the two-dimensional growth, the main driving force is the decrease in the total energy of grain boundaries on the surface ($F_{Gb,2-d}$).

$$\Delta F_{Gb,2-d} = \frac{2}{3} \Delta F_{Gb,2-d} = \frac{2a * \gamma_{Gb}}{d}, \tag{2}$$

where $\gamma_{GB}$ – grain boundary energy
d – grain diameter
a – geometry factor of order 1

2) For the three-dimensional growth, the main driving force is the decrease in the total energy of grain boundaries by volume ($F_{Gb,3-d}$).

$$\Delta F_{Gb,3-d} = \frac{3a * \gamma_{Gb}}{d}, \tag{2}$$

Other driving forces are the following:

3) Anisotropy of surface/interface energy ($F_{S/i}$). It occurs during the decrease in surface energy and elastic energyЖ

$$\Delta F_{S/I} = \frac{\Delta \gamma}{h}, \tag{3}$$

where $\Delta\gamma = \gamma_{gb} - \gamma_s$ is difference of average surface and interface energy between the two textures (film – substrate boundary)
h – film thickness.

4) An anisotropy of strain energy. It occurs during a substrate temperature change between the film deposition and crystallization moment. It is caused by the film/substrate expansion coefficient mismatch

$$\Delta F_e = \Delta M (\Delta \alpha \Delta T)^2, \tag{4}$$

where ΔM - difference in biaxial modulus between the two textures (depends on direction),
Δα – difference between the thermal expansion coefficients of film and substrate,
ΔT – difference between deposition and annealing temperatures.

In addition, forces leading to stagnation appear during a film crystallization. The force balance for surface and boundary energies leads to the formation of ''grooves'' at the intersection of grain boundaries with the upper and lower surfaces [46] (See Supplementary Section 1). It corresponds to the ''specimen thickness'' effect which appears upon grain size exceeding thickness by the factor from one to three [56, 57]. The angle at the bottom of the groove is determined by the relative surface energy of the free surface and the grain boundary. This angle determines the force required to pull the grain boundary out of the groove. According to the calculations, the force value is determined by the angle at the grain boundary, which in its turn, determines the critical radius of the catenoid formed at the grain boundary [53, 54].

A catenoid curvature during stagnation is determined as

$$\kappa_{crit} = \frac{\gamma_{gb}}{\gamma_s * h}, \qquad (5)$$

It also could be calculated using Johnson – Mehl structure square value ($A_0$) [54]:

$$\kappa_{crit} = \frac{\gamma_{gb}}{\gamma_s * h}, \qquad (6)$$

Johnson – Mehl structure square value could also be used for stagnation force calculation:

$$F_{stag} = \frac{a}{\sqrt{A_0}}, \qquad (7)$$

For calculations, the authors use a=0,4 value [54], when the stagnation force is given by:

$$F_{stag} = \frac{0,4}{\sqrt{A_0}} \qquad (8)$$

Another known stagnation force is the triple junction drag [58-64]. This force is applied at the triple junction point. And the last factor, which causing stagnation is solute drags. In most cases, the presence of dissolved drags and impurities in the material and their accumulation at the boundary inhibits the movement of grain boundaries. In our case, the deposition is performed in high vacuum using ultrapure material (99.999%), so we do not take into account the influence of impurities to the described process. All the mentioned forces were calculated for the observed metal – substrate system (see Supplementary Table S5).

Of interest to us is abnormal growth, when grain size is significantly larger than its thickness [65-67]. An abnormal growth is driven by a higher driving force for some grains, which is explained by different interface energies for different crystallographic orientations [66]. The trade-off between minimization of stresses from the film thickness and stress from the interface leads to the growth of different orientations. The best stress compensation in face-centered cubic metal films occurs for the (001) orientation, while the best surface energy compensation occurs for the (111) orientation. After deposition, one can see increase the fraction of (111) orientation and in the ideal case of no stagnation, only the (111) orientation is observed at the end of the film growth. Typically, for a real system a grain growth is limited by the intrinsic stress accumulated in the thin film after the deposition. According to equations (1), (2), (3), (4), (8) stress is the only growth force dependent on external parameters and thus could be used for grain growth control. At the first stage, most of the grains are still in the elastic regime and the (001) grains have the growth advantage due to their low strain energy density. At the next stage, the growth of (111) grains occurs until the stagnation moment. After stagnation, an abnormal growth continues, its conditions are determined by the film stress value, which stimulates the (001) grains growth, contributing to the average grain size increase (See Supplementary Section 3). The stress value is determined by the deposition parameters (see Supplementary Table S2) [54]. We found the thickness/temperature ratio providing the maximum stress and thereby maximum grain size during recrystallization: "Golden rectangle" (Fig. 2c).

We carried out a complete characterization of structural, morphological and optical properties of the fabricated samples. The analysis of the film with the best parameters is shown below (Fig. 4). High-resolution scanning electron microscopy (HR-SEM) was done to visualize film surface and estimate grain size (Fig. 4a). We used atomic force microscopy to measure films morphology and surface roughness). The best sample root mean square (RMS) roughness was less than 1.5 nm measured over 2.5×2.5 µm² area (Fig. 4b). Electron backscatter diffraction (EBSD) is measured to analyze the domain structures and extract the average grain size of films. The best sample mean crystalline size higher than 300 nm measured over 15×15 µm² area (Figs 4d,e). To measure optical properties a multi-angle spectroscopic ellipsometry was used. We observe the dominating contribution of grain boundaries to dielectric permittivity (Fig. 4c, f), the real part becomes more negative with the increasing grains size (Fig. 4c) indicating higher conductivity. The polycrystalline film with a great number of small grains has the inferior $\varepsilon_1$ compared to JC and PCBG film. Observed decrease in negative $\varepsilon_1$ (conductivity) is primarily due to the increased number of structural defects (including grain boundaries) in the films leading to the increased electron-phonon interactions, which make the films less metallic. In general, the same influence of the films grain size on $\varepsilon_2$ is observed (Fig. 4f).

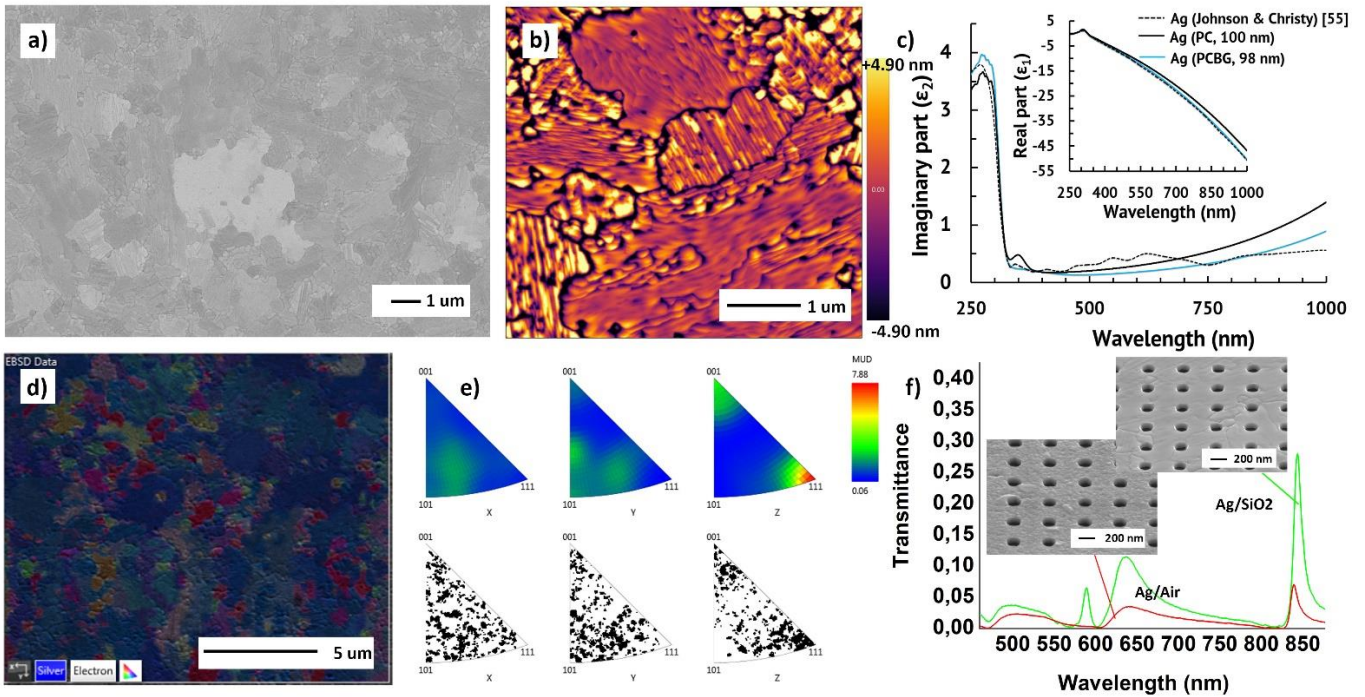

**Figure 4.** Thin films characterization: a) SEM image (7000x magnification) of 100 nm-thick polycrystalline big grain silver thin films on a quartz substrate; b) AFM scan of Ag film with an RMS roughness of less than 1.5 nm; c) real and imaginary parts of Ag films, reported be Jonson & Christy [55] and the polycrystalline silver thin films permittivity, measured by multi-angle spectroscopic ellipsometry; d, e) orientation map of Ag film grains, obtained using with EBSD analysis; f) Zero-order transmission spectra of these films for polycrystalline (red) and PCBG (green) single-crystalline silver films, insets: SEM images (40000x magnification) of 100 nm-thick silver films and structural features at their nanostructuring by the e-beam lithography

In order to estimate the plasmonic performance of the PCBG silver films, we compared the resonance quality of the perforated structure with different silver crystalline structure (See Methods section). Figure 4f is a SEM illustration of resonant nanostructures fabricated using 100 nm-thick silver films with polycrystalline, PCBG and single-crystalline structure. Zero-order transmission spectra of these devices are shown in plot , illustrating the peaks related to the Fano resonances that can vary in intensity by one order of magnitude (see λ ≈ 850 nm). Actually, the RMS surface roughness of higher than 1.5 nm and nanocrystalline structure of polycrystalline (red) sample causes a weaker light coupling to plasmonic surface wave at the silver/air interface and even absence of the resonance from the spectrum–see curve at λ ≈ 580 nm. The structural perfection and high optical quality factor of nanostructures are observed for the high quality deposited PCBG (green) silver films due to their big grains structure.

Finally, we fabricated plasmonic devices based on the developed deposition technology. A fabricated plasmonic laser cavity is characterized by an ultra-low loss (40 $cm^{-1}$) owing to the use of an ultra-high-quality silver film and the hybrid plasmonic-photonic mode. Lasing at 628 nm, with a linewidth of 1.7 nm and a directivity of 1.3° is observed [18]. A new type of plasmonic sensor was developed based on the use of the Ebbesen effect of the extraordinary transmission (EOT) of light. A record level of sensitivity in plasmonics was demonstrated, pawing a way to ultrasensitive sensors for biomarkers molecules using plasmonic structure. The minimum recorded dye molecule concentration was 20 pg/ml (3 ppt) [19].

In conclusion, we emphasized the need for high quality thin films beyond the single-crystalline metals and formulated the requirements to the desired film properties. We performed the review on the main results and challenges in deposition of high quality polycrystalline thin films. Following both the theoretical considerations and experimental evidence, we proved that strain is the only growth force that is dependent on external parameters. This enabled development of the three-step approach for e-beam evaporation of polycrystalline big grain silver films: 1) deposition of nanocrystalline film with sufficient stored internal energy, 2) self-crystallization and stored energy relaxation, 3) final crystallization at well-tuned annealing. We observed that the both crystallization processes are controlled by the evolutionary selection growth, whereas the key feature for its optimization is the use of the well-defined ratio of deposition parameters and a subsequent heat treatment for the precise control of the resulting stresses. We fabricated the optimized 100 nm-thick silver films, which exhibit the nm-scale surface roughness and average grain size bigger than 0.5 μm, occupying >80% of the entire film surface. These films features enable low optical losses and fabrication of ultra-high-quality plasmonic devices. The resonance quality of the perforated structure for silver thin films with different crystalline structure was compared. The structural perfection, low roughness and high optical quality factor of nanostructures are evident primarily for the quality of deposited PCBG silver films – their big grains structure. We believe that the PCBG process could be used for deposition of various high quality metal thin films and could be easily integrated in a planar top-down device fabrication technology.

**Methods**
**Preparation of epitaxial films.** Silver thin films were deposited on prime-grade Siegert silica substrates using 10 kW e-beam evaporator (Angstrom Engineering) with a base pressure lower than $3 \times 10^{-8}$ Torr. We first cleaned the wafers in a 2:1 sulfuric acid: hydrogen peroxide solution (80 °C), followed by further cleaning in isopropanol to eliminate organics. After cleaning, we immediately transferred the wafers into the evaporation tool and pumped the system down to limit native oxide growth. All films were grown using 5N (99.999%) pure silver. Films were deposited with rate of 0.5–10 A·s$^{-1}$ measured with quartz monitor at approximate source to substrate distance of 30 cm. Deposition is done in three steps using PCBG process.

**Scanning electron microscopy.** In order to check the quality and uniformity of the deposited layers silver films surfaces after deposition were investigated by means of a scanning electron microscope. All SEM images were obtained using in-lens detector and the accelerating voltage 5 kV and working distance from the sample to detector from 1 to 4 mm. Magnifications 3000, 7000, 15000 and 50000 were used to fully analyze samples.

**Atomic force microscopy.** The atomic force microscope Asylum Jupiter with probe with nominal tip radius of 2 nm was used. AFM images were obtained by using Tapping mode and the scanned area was 2.5×2.5 μm$^2$.

**Electron back scattered diffraction (EBSD) characterisation.** The Ag films were observed and structurally characterized by field emission scanning electron microscopy. The crystal orientation maps of the Ag films were obtained by FE-SEM equipped with an EBSD system. EBSD patterns were acquired at the following shooting modes: tilt angle – 70°, accelerating voltage – 10 keV, probe current – 1.7 nA and scan sizes 2 × 2 μm$^2$ and 20 × 20 μm$^2$ for SC film. EBSD has proven to be a useful tool for characterizing the crystallographic orientation aspects of microstructures at length scales ranging from dozens of nanometers to millimeters in the scanning electron microscope. Detector provides single grains detection by means of orientation measurement based on acquired Kikuchi patterns. Colored image represents grains orientation map, correlation between colors and orientations is shown on triangle diagram at the bottom left corner. We extract average grain size for our polycrystalline films using the embedded software package for an EBSD image processing AZtecHKL software package. We extract average grain size for our polycrystalline films using the embedded software package for an EBSD image processing. The NC film was e-beam evaporated onto a liquid nitrogen cooled quartz substrate, the conditions were adjusted so that the film had an average grains size around 20 nm.

**Ellipsometry.** Dielectric functions of the silver films were measured using a multi-angle spectroscopic ellipsometer (SER 800, Sentech GmbH). We specifically measured the optical constants of the substrates used in the deposition process to eliminate any discrepancy and uncertainty. These measured substrate optical constants and silver thickness are fixed in the subsequent data fitting for all samples, and only the silver parameters are allowed to change. Modeling and analysis were performed with the ellipsometer SENresearch 4.0 software. Measurement spectral wavelength range was from 240 to 1000 nm, with an interval of approx. 2 nm, and the reflected light was analyzed at incidence angles of 50°, 60°, 70°. To characterize the optical losses, the real ($\varepsilon_1$) and imaginary ($\varepsilon_2$) parts of the dielectric permittivity were extracted by fitting the measured raw ellipsometric data (Ψ and Δ). In our fitting, we used a bilayer Ag/Si structural model and a simple phenomenological Brendel-Bormann (BB) oscillator model to interpret both the free electron and the interband parts of the dielectric response of our samples:

$$\hat{\varepsilon}(\omega) = \varepsilon_\infty - \frac{\omega_p^2}{\omega^2 + i\Gamma_D \omega} + \sum_{j=1}^{k} \chi_j(\omega), \qquad (1)$$

where ωp is the plasma frequency, $\varepsilon_\infty$ is the background dielectric constant, $\Gamma_D$ is Drude damping, $\chi_j(\omega)$ is BB oscillators interband part of dielectric function, and k is the number of BB oscillators used to interpret the interband part of the spectrum.

**Transmission spectra measurement.** The properties of transmission spectra were studied by the Ntegra Spectra (NT-MDT, Russia) multichannel microscope&spectrometer. To obtain transmission spectra at normal incidence, a collimated beam from a white light source (a Nikon illuminator, 75 W) was prepared and reduced to a diameter of 300 μm by an aperture located in front of the chip under study; the beam divergence angle of the beam is less than 0.1°.

## Acknowledgements

Samples were made and measured at the BMSTU Nanofabrication Facility (FMN Laboratory, FMNS REC, ID 74300).

## Author contributions statement

A.S.B., I.A.R. (Ilya A. Ryzhikov) and I.R. (Ilya A. Rodionov) conceptualized the ideas of the project. A.S.B., D.O.M, E.S.L. fabricated thin films. O.S.S., D.A.B. performed morphology characterization. K.A.B conducted characterization of the thin films optical parameters. G. M. Y. and A. V. B. conducted resonance research. S.S.A. processed data. I.R. reviewed and edited the manuscript. I.R. supervised the project. All authors analyzed the data and contributed to writing the manuscript.

## Additional information

The authors declare no conflict of interest.